\documentstyle[12pt]{article}
\def\hybrid{\topmargin 0pt      \oddsidemargin 0pt
	\headheight 0pt \headsep 0pt
	\textheight 9in         % US paper
	\textwidth 6.25in       % A4 paper
	\marginparwidth .875in
	\parskip 5pt plus 1pt   \jot = 1.5ex}

\catcode`\@=11
\def\marginnote#1{}
\newcount\hour
\newcount\minute
\newtoks\amorpm
\hour=\time\divide\hour by60
\minute=\time{\multiply\hour by60 \global\advance\minute by-\hour}
\edef\standardtime{{\ifnum\hour<12 \global\amorpm={am}%
	\else\global\amorpm={pm}\advance\hour by-12 \fi
	\ifnum\hour=0 \hour=12 \fi
	\number\hour:\ifnum\minute<10 0\fi\number\minute\the\amorpm}}
\edef\militarytime{\number\hour:\ifnum\minute<10 0\fi\number\minute}

\def\draftlabel#1{{\@bsphack\if@filesw {\let\thepage\relax
   \xdef\@gtempa{\write\@auxout{\string
      \newlabel{#1}{{\@currentlabel}{\thepage}}}}}\@gtempa
   \if@nobreak \ifvmode\nobreak\fi\fi\fi\@esphack}
	\gdef\@eqnlabel{#1}}
\def\@eqnlabel{}
\def\@vacuum{}
\def\draftmarginnote#1{\marginpar{\raggedright\scriptsize\tt#1}}

\def\draft{\oddsidemargin -.5truein
	\def\@oddfoot{\sl preliminary draft \hfil
	\rm\thepage\hfil\sl\today\quad\militarytime}
	\let\@evenfoot\@oddfoot \overfullrule 3pt
	\let\label=\draftlabel
	\let\marginnote=\draftmarginnote
   \def\@eqnnum{(\theequation)\rlap{\kern\marginparsep\tt\@eqnlabel}%
\global\let\@eqnlabel\@vacuum}  }

\def\numberbysection{\@addtoreset{equation}{section}
	\def\theequation{\thesection.\arabic{equation}}}

\def\underline#1{\relax\ifmmode\@@underline#1\else
	$\@@underline{\hbox{#1}}$\relax\fi}

\def\titlepage{\@restonecolfalse\if@twocolumn\@restonecoltrue\onecolumn
     \else \newpage \fi \thispagestyle{empty}\c@page\z@
	\def\thefootnote{\fnsymbol{footnote}} }

\def\endtitlepage{\if@restonecol\twocolumn \else  \fi
	\def\thefootnote{\arabic{footnote}}
	\setcounter{footnote}{0}}  %\c@footnote\z@ }
\catcode`@=12
\relax
\def\ie{\hbox{\it i.e. }}

\def\ve{\varepsilon}
\def\nn{\nonumber}
\def\beq{\begin{equation}}
\def\eeq{\end{equation}}
\def\bea{\begin{eqnarray}}
\def\eea{\end{eqnarray}}
\def\ep{\epsilon}
\def\si{\sigma}
\relax
\def\half{{1\over 2}\;}

\numberbysection
\hybrid
\begin{document}
\begin{titlepage}
\begin{center}
December~1994 \hfill    PAR--LPTHE 94/41 \\[.5in]
{\large\bf Renormalisation group calculation of correlation
functions for the 2D random bond Ising and Potts models}\\[.5in]
        {\bf Vladimir Dotsenko\footnote{Also at the Landau Institute for
Theoretical Physics, Moscow}, Marco Picco and Pierre
Pujol} \\
	{\it LPTHE\/}\footnote{Laboratoire associ\'e No. 280 au CNRS}\\
       \it  Universit\'e Pierre et Marie Curie, PARIS VI\\
       \it Universit\'e Denis Diderot, PARIS VII\\
	Boite 126, Tour 16, 1$^{\it er}$ \'etage \\
	4 place Jussieu\\
	F-75252 Paris CEDEX 05, FRANCE\\
	dotsenko,picco,pujol@lpthe.jussieu.fr
\end{center}

\vskip .5in
\centerline{\bf ABSTRACT}
\begin{quotation}
We find the cross-over behavior for the spin-spin
correlation function for the 2D Ising and 3-states Potts model with random
bonds at the critical point. The procedure employed is the renormalisation
approach of the perturbation series around the conformal field theories
representing the pure models.
We obtain a crossover in the amplitude for the correlation function for
the Ising model which doesn't change the critical exponent, and a shift in
the critical exponent  produced by randomness in the case of the Potts
model. A comparison with numerical data is discussed briefly.

\end{quotation}
\end{titlepage}
\newpage

\section{Introduction}

In the studies of critical phenomena in real physical systems, impurities
and inhomogeneities are always present. Many theoretical models have been
proposed for the study of random models. In most of these cases the replica
trick for quenched systems is employed. This corresponds to computing the
averaged free energy by taking $n$ copies of the system and going to the
limit $n\rightarrow0$. For
models with random bonds, the main problem is to determine if the
randomness leaves unchanged the critical properties of the pure system or
if the singularities of the thermodynamical functions are eliminated.
First results, which suggest an intermediate situation, have been obtained
by Harris
and Lubensky \cite{harrlub}, Grinstein and Luther \cite{grins} and
Khmelnitskii \cite{khmel} using the standard $\phi^4$ theory. Other cases,
like long-range correlated quenched defects have also been considered, (see
for example \cite{weinrib,korzh}.) A first step in the understanding of the
relevance of randomness was given by the Harris criterion \cite{harris}:
the randomness is relevant (irrelevant) if the specific heat exponent of
the pure model is positive (negative). Two dimensional systems are
particularly interesting because of the rich structure of conformal
invariance in this dimension. Assuming that a random model has a critical
point with second order phase transition, the main interesting problem is
to determine which conformal field theory represents this model at the
infrared fixed point. Let us also mention that an exact result has been
obtained by McCoy and Wu \cite{mccoy} who considered a two-dimensional
Ising model where only vertical bonds on a square lattice were allowed to
acquire the same random value. They found that the logarithmic singularity
of the specific heat disappeared completely.

The models that we will study in this paper are the two dimensional Ising
and Potts models with random bonds. For the case of Ising model, Harris
criterion doesn't provide a qualitative
answer of the relevance of the randomness (because the specific heat
exponent for the $2-D$ Ising model is $0$). First results were obtained by
Dotsenko and Dotsenko \cite{dots1} who showed that near the critical point,
this model can be represented by an $n=0$ Gross-Neveu model \cite{GN}. With
this technique, they found that the specific heat singularity get smoothed
as $ln(ln({1\over|t|}))$ where $|t|$ is the reduced
temperature. Calculation of spin-spin correlation function by this
technique which involves non-local fermionic representation of $<\sigma
\sigma>$, was later questioned by Shalaev and Shankar
\cite{shalaev,shankar} who gave arguments that the asymptotic behavior of
this correlation function is unchanged by the randomness (see also
\cite{ludwig2}.) Some arguments on why computations using non-local
fermionic representation fail to give the correct result for the spin-spin
correlation function in the random case will be given at the end of the
paper.
More recently, this situation was questioned by Ziegler
\cite{ziegler}. This author claims that non-perturbative effects introduce
an intermediate phase around the critical point of the pure model. On the
other hand, numerical simulations of the Ising model \cite{talapov,adsw}
seem to confirm the theoretical predictions of the specific heat and
spin-spin correlation function asymptotic behavior. For the Potts model,
Harris criterion predicts that randomness is relevant and changes the
critical behavior.
Using conformal field theory techniques, Ludwig \cite{ludwig1} also
perturbatively computed  a shift in the critical exponent of the energy
operator in the case of the random Potts model (this critical exponent for
the Ising model is also unchanged).

The paper, which complete the work presented in \cite{us}, is organized as
follows. In section $2$ we explain how the replica trick method
can be used to deal with the partition function of the $2 D$ Ising and
Potts models with random bonds. Near the critical point, these models
are represented by perturbed conformal field theories. We give a brief
summary of the Coulomb Gas representation
of conformal field theories needed to pertubatively compute correlation
functions. We recall also a kind of ``$\epsilon$'' regularization which
consists of shifting the central charge of the pure model. Then, in section
$3$, we sketch the renormalisation group equations for the coupling
constant and correlation function. Specializing then to the particular case
of the Ising model, we find a cross-over in the amplitude at intermediate
distances produced by randomness. In the case of the Potts
model, we show that the spin exponent is modified in a power
series of $\epsilon$ due to the randomness. Finally, section $4$ contains a
summary of our results and a discussion. Technical details of the
renormalisation group calculation and the calculation of integrals are put
into Appendices.

\section{The model}

Our starting point is the 2-D Ising-Potts Hamiltonian at the critical
temperature:
\beq
S_0 =\displaystyle\sum_{<i,j>} (\beta J_0) \delta_{s_i,s_j}
\eeq
where $J_0$ is the coupling between spins, $s_i$ are the spin
variables and $<i,j>$ means that the sum is restricted to nearest neighbor
spins. The unperturbed partition function is:
$$
Z_0 = Tr_{s_i} e^{-S_0}
$$
Here, $ Tr_{s_i}$ means a summation over all spin configurations. The
addition of a position dependent random coupling constant produces the
following change in the partition function: $ J_0 \to J_0 +\Delta
J_{i,j}$. It is well known that at the critical point the unperturbed
models are represented in the continuum by unitary minimal conformal field
theories in which the relevant operators are the spin operator $\sigma$
corresponding to $s_i$ and the energy operator $\varepsilon$ corresponding
to $\delta_{s_i,s_j}$. In the continuum the term $\Delta J_{i,j}$ can be
written as a position dependent
mass term $m(z)$, which gives the total partition function in the form~:
\beq Z =
Tr_{s_i} e^{-S_0-\int m(z) \varepsilon(z) d^2 z}
\eeq
In order to compute
the free energy, we use the replica method. Taking the partition function
of $n$ identical copies of the system and analytically continuing to the
limit $n\to0$ gives the quenched free energy
$$
-\beta \overline{F} =
\overline{\ln(Z)} = \displaystyle\lim_{n\to0}{\overline{{Z^n - 1 \over n}}}
$$
where:
\beq
Z^n = \prod_{a=1}^n  Tr_{a,s_i}
e^{-\displaystyle\sum_{a=1}^n S_{0,a} - \int m(z)\displaystyle\sum_{a=1}^n
\varepsilon_a(z) d^2z}
\eeq
the average of $ Z^n$ is made with a Gaussian distribution for $m(z)$:
$$
\overline{Z^n} = \int \prod_z dm(z) Z^n e^{-{1\over2g_0}(m(z)-m_0)^2}
$$
which gives:
\beq
\label{z}
\overline{Z^n} = \prod_{a=1}^n Tr_{a,s_i} e^{-\displaystyle\sum_{a=1}^n
S_{0,a} + g_0 \int \displaystyle\sum_{a,b=1}^n
\varepsilon_a(z)\varepsilon_b(z) d^2z - m_0 \int \displaystyle\sum_{a=1}^n
\varepsilon_a(z) d^2z}
\eeq
The average with a more complicated
distribution gives higher order cumulants, producing in (\ref{z}) terms
which can be shown to be irrelevant by power counting \cite{ludwig1}. Also,
the terms in
$\int \displaystyle\sum_{a,b=1}^n \varepsilon_a(z)\varepsilon_b(z) d^2z$
containing the same replica label produce irrelevant operators and can be
omitted. We obtain then the following partition function:
\beq
\prod_{a=1}^n Tr_{a,s_i} e^{-\displaystyle\sum_{a=1}^n S_{0,a} + g_0
\int \displaystyle\sum_{a\not=b} \varepsilon_a(z)\varepsilon_b(z) d^2z -
m_0 \int \displaystyle\sum_{a=1}^n \varepsilon_a(z) d^2z}
\eeq
In the limit
$ m_0\rightarrow0$, this model corresponds to a conformal field theory
perturbed by the term that is quadratic in the $\varepsilon$
operator. Then the ``evolution'' of the coupling constants $g_0$ and $m_0$
under a renormalisation group (R.G.) transformation can be analyzed as well
as the behavior of the correlation functions.  In the calculation of
correlation
functions $<O(0)O(R)>$, where $O$ is some local operator, we will proceed
perturbatively:
$$
<O(0)O(R)> = <O(0)O(R)>_0+<S_IO(0)O(R)>_0+{1\over2}<S_I^2O(0)O(R)>_0+...
$$
where $<>_0$ means the expectation value taken with respect to $S_0$ and
\beq
S_I =
\int H_I(z) d^2z =  g_0 \int \displaystyle\sum_{a\not=b}
\varepsilon_a(z)\varepsilon_b(z) d^2z
\eeq
The operator $O$ is then renormalised as
$$
O\rightarrow O ( 1 + A_1g_0 + A_2g_0^2 + A_3 g_0^3 + \cdots ) \equiv
Z_0 O
$$
The integrals of correlation functions involved in the calculation can be
performed by analytic continuation with the Coulomb-gas representation of a
conformal field theory \cite{dots2} where the central charge is $c = {1
\over 2} +
\epsilon'$. The $\epsilon'$ term corresponds to a short distance regulator
for the integrals. In addition, we also used an infrared (I.R.) cut-off
$r$. The
result is then expressed as an $\epsilon'$ series with coefficients
depending on $r$. The limit $\epsilon' \rightarrow 0$ corresponds to
the pure Ising
model at the critical point while the Potts model is obtained for some
finite value of
$\epsilon'$. We recall here some notations of the Coulomb-gas representation
for the vertex operators \cite{dots2}. The central charge $c$ will be
characterized in the following by the parameter $\alpha^2_+ = {2p\over2p-1}
={4\over 3} + \ep $ with
\bea
\label{c}
c=1-24 \alpha_0^2 \hskip 0.5cm &;& \hskip 0.5cm \alpha_\pm = \alpha_0 \pm
\sqrt{\alpha_0^2 +1} \\
\alpha_+ \alpha_- &=& -1 \nonumber
\eea
Note that for the pure 2D Ising model
 $\alpha_+^2 ={4\over 3} $ and $c = {1\over 2}$
while  for the $3$-state Potts model  $\alpha_+^2 =
{6\over 5}$, $c={4\over 5}$ and $\epsilon = -{2\over 15}$.
 The vertex operators are defined by
\beq
V_{nm}(x) = e^{i\alpha_{nm} \phi(x)}
\eeq
where $\phi(x)$ is a free scalar field and where the $\alpha_{nm}$ are
given by
\beq
\alpha_{nm}=\half (1-n) \alpha_- + \half (1-m) \alpha_+
\eeq
The conformal dimension of an operator $V_{nm}(x)$ is
$\Delta_{nm} =
-\alpha_{\overline{nm}} \alpha_{nm}$ with
\beq
\label{al}
\alpha_{\overline{nm}} = 2\alpha_0 -\alpha_{nm} =\half (1+n) \alpha_- +
\half (1+m) \alpha_+
\eeq
The spin field $\sigma$ can be represented by the vertex operator
$V_{p,p-1}$ whereas $V_{1,2}$ corresponds to the energy operator
$\varepsilon$. In the same way, we associate $ e^{i\alpha_+ \phi(x)}$ to
the screening charge operator $ V_+(x)$.
Note that in the Ising case the $\sigma$ operator can also be represented
by  the $V_{21}$ operator (since both operators coincide in the limit
$\epsilon \rightarrow 0$). So, we can represent our spin operator by
$V_{k,k-1}$ where $k = {2 + 3 \lambda  \epsilon \over 1 + 3
\epsilon}$. We have $\lambda = 2$ for $V_{21}$ and  $\lambda = {1\over2}$
for $V_{p,p-1}$.
\section{Renormalisation Group Equations}
In this section, we will deal with the computation of correlation functions
of operators $\varepsilon$ and $\sigma$. To compute them, one needs to
determine the effect of the random
coupling on the operators $\varepsilon$ and $\sigma$ and compute the
renormalised operators $\varepsilon '$ and $\sigma '$. This means that we
want to compute the functions $Z_\varepsilon$ and $Z_\sigma$ such that
\beq
\varepsilon ' = Z_\varepsilon \varepsilon \quad \mbox{and} \quad
\sigma ' = Z_\sigma \sigma
\eeq
A convenient way to define $Z_\varepsilon$ and $Z_\sigma$ is to
consider the more general action
\beq
\displaystyle\sum_{a=1}^n S_{0,a}
 -g_0 \int \displaystyle\sum_{a,b=1}^n \varepsilon_a(z)\varepsilon_b(z)
d^2z + m_0 \int \displaystyle\sum_{a=1}^n \varepsilon_a(z) d^2z - h_0 \int
\displaystyle\sum_{a=1}^n \sigma_a(z) d^2z
\eeq
This merely corresponds to the action used in (\ref{z}) with an additional
coupling of the $\sigma$ field. Then, with the help of the operator algebra
(O.A.) coming from contractions between $\ve$ and $\si$ operators, we will
compute the effect of the $g_0
\int \displaystyle\sum_{a,b=1}^n \varepsilon_a(z)\varepsilon_b(z) d^2z $
term (\ie the random part of the model) on the coupling terms $m_0 \int
\displaystyle\sum_{a=1}^n \varepsilon_a(z) d^2z$ and $h_0 \int
\displaystyle\sum_{a=1}^n \sigma_a(z) d^2z$. More precisely, we will
compute
\beq
\sum_i \left( \left(g_0 \int \displaystyle\sum_{a,b=1}^n
\varepsilon_a(z)\varepsilon_b(z) d^2z \right)^i m_0 \int
\displaystyle\sum_{a=1}^n \varepsilon_a(z) d^2z \right) \simeq m \int
\displaystyle\sum_{a=1}^n \varepsilon_a (z) d^2z
\eeq
and
\beq
\sum_i \left( \left(g_0 \int \displaystyle\sum_{a,b=1}^n
\varepsilon_a(z)\varepsilon_b(z) d^2z \right)^i h_0 \int
\displaystyle\sum_{a=1}^n \sigma_a(z) d^2z \right) \simeq h \int
\displaystyle\sum_{a=1}^n \sigma_a (z) d^2z
\eeq
$m$ and $h$ being the renormalised coupling constants. Obviously, this
computation will be perturbatively made only up
to some finite power in $g_0$. In fact, the first step of the computation
will be to determine the renormalised $g$
constant on which $Z_\varepsilon$ and $Z_\sigma$ depend.
\subsection{Renormalisation of the coupling constant $g$.}
The renormalisation of the coupling constant $g$ will be determined
directly by a perturbative computation. $g$ is also
given by the O.A. producing
\bea
g_0 \int \displaystyle\sum_{a,b=1}^n \varepsilon_a(z)\varepsilon_b(z) d^2z
&+& {1\over 2} \left(g_0 \int \displaystyle\sum_{a,b=1}^n
\varepsilon_a(z)\varepsilon_b(z) d^2z\right)^2 +  \\
&+& {1\over 6} \left(g_0 \int
\displaystyle\sum_{a,b=1}^n \varepsilon_a(z)\varepsilon_b(z) d^2z\right)^3
+ \cdots =  g \int \displaystyle\sum_{a,b=1}^n
\varepsilon_a(z)\varepsilon_b(z) d^2z \nonumber
\eea
with $g=g_0 + A_2 g_0^2 + A_3 g_0^3 + \cdots $ where $A_2$ comes from
\beq
{1\over 2} \int \displaystyle\sum_{a,b=1}^n
\varepsilon_a(z)\varepsilon_b(z) d^2z \int \displaystyle\sum_{a,b=1}^n
\varepsilon_a(z)\varepsilon_b(z) d^2z = A_2 \int
\displaystyle\sum_{a,b=1}^n \varepsilon_a(z)\varepsilon_b(z) d^2z + \cdots
\eeq
and $A_3$ from
\bea
{1\over 6} \int \displaystyle\sum_{a,b=1}^n
\varepsilon_a(z)\varepsilon_b(z) d^2z && \!\!\!\!\!\!\int
\displaystyle\sum_{a,b=1}^n
\varepsilon_a(z)\varepsilon_b(z) d^2z \int \displaystyle\sum_{a,b=1}^n
\varepsilon_a(z)\varepsilon_b(z) d^2z \nonumber \\
&&= A_3 \int \displaystyle\sum_{a,b=1}^n
\varepsilon_a(z)\varepsilon_b(z) d^2z + \cdots
\eea
Here, we will only perform the computations up to the third order in
$g_0$. The technical details of the computations of $A_2$ and
$A_3$ are given in the appendix A with the following result for
$g(r)$~:
\beq
g(r) = r^{-3\epsilon}\left( g_0 - g_0^2  4\pi (n-2){r^{-3\epsilon}\over
3\epsilon} + g_0^3 8 \pi^2 (n-2) {r^{-6\epsilon}\over
3\epsilon} \left(1+ {2 (n-2)\over 3\epsilon}\right)\right)
\eeq
Note that we multiply the result by $r^{-3\epsilon}$ in order to obtain
a dimensionless coupling constant $g(r)$.

In the computation of correlation functions, we will need the
$\beta$-function associated to the renormalisation group equations of
$g(r)$. It can be derived directly from the previous
expression of $g(r)$ with the result~:
\beq
\beta(g) = {dg\over dln(r)} = -3\epsilon g(r) + 4 \pi (n-2) g^2(r) -16
\pi^2 (n-2) g^3(r) + O(g^4(r))
\eeq
Finally, taking the limit $n \rightarrow 0$, we obtain for the
$\beta$-function up to the third order~:
\beq
\label{tobf}
\beta(g) =  -3\epsilon g - 8 \pi g^2 + 32 \pi^2 g^3
\eeq
We can then immediately note that, in the limit $ \epsilon \rightarrow 0$
(\ie the random Ising model), $\beta(g)$ has an infrared fixed point
at $g=0$.
In the case where $\epsilon < 0$ (\ie the random Potts
model) the infrared fixed point is located at $g_c = -{3\epsilon \over 8
\pi} + {9\epsilon^2 \over 16\pi} + O(\epsilon^3) $.

\subsection{Renormalisation of $\sigma$ and $\varepsilon$}

In order to be able to compute the correlation functions of $\sigma$ and
$\varepsilon$, the second step is to determine the effect of the
renormalisation on these operators. One needs to compute the
multiplicative functions $Z_\sigma$ and $Z_\varepsilon$. This will be made
by computing the renormalised coupling terms $m\int
\displaystyle\sum_{a=1}^n \varepsilon_a (z) d^2z = (m_0 Z_\varepsilon )\int
\displaystyle\sum_{a=1}^n  \varepsilon_a(z) d^2z $ and
$h\int \displaystyle\sum_{a=1}^n \sigma_a (z) d^2z = (h_0
Z_\sigma)\int \displaystyle\sum_{a=1}^n \sigma_a(z) d^2z $.
A direct computation of both $m$ and $h$ will provide
us with the functions $Z_\sigma$ and $Z_\varepsilon$. As for the
computation of $g=Z_g g_0$, we will compute in perturbation~:
\bea
m_0\int\sum_{a=1}^n\ve_a(z) d^2z && \!\!\!\!\!\! + g_0 m_0 \int
\sum_{a,b=1}^n \ve_a(z)\ve_b(z) d^2z \int\sum_{a=1}^n\ve_a(z) d^2z +
\\&& +
{g_0^2\over2}m_0 \left(\int \sum_{a,b=1}^n \ve_a(z)\ve_b(z)\right)^2
\int\sum_{a=1}^n\varepsilon_a(z) d^2z  + \cdots
= m\int\sum_{a=1}^n\ve_a(z) d^2z\nn
\eea
and $m=m_0(1+B_1 g_0 + B_2 g_0^2 + \cdots)$ with $B_1$ defined by
$$
\int \sum_{a,b=1}^n \ve_a(z)\ve_b(z) d^2z \int\sum_{a=1}^n\ve_a(z) d^2z =
B_1 \int\sum_{a=1}^n\ve_a(z) d^2z
$$
and $B_2$ by
$$
\left(\int \sum_{a,b=1}^n \ve_a(z)\ve_b(z)\right)^2
\int\sum_{a=1}^n\varepsilon_a(z) d^2z = B_2 \int\sum_{a=1}^n\ve_a(z) d^2z
$$
The details of the computation are presented in the appendix B, with the
result
$$r^{-1+{3\over2} \epsilon}
 m(r) = m_0 \Bigl( 1-4\pi
(n-1)g_0 {r^{-3\epsilon}\over
3\epsilon} + 4 \pi^2 (n-1)g_0^2{r^{-6\epsilon}\over
3\epsilon} (1+{4n-6\over 3\epsilon} )\Bigr)$$
Here again, we multiply $m(r)$ by $r^{-1+{3\over2} \epsilon}$ in order to
obtain a dimensionless coupling constant. We also give the R.G. equation
for $Z_\ve$, which we will need later.
\beq
\label{eqzve}
{d ln(Z_\ve(r)) \over d ln(r)} =   4\pi (n-1) g
 -   8 \pi^2 (n-1) g^2
\eeq
Similarly, for the coupling constant $h_0$, we compute up to the third
order~:
$$
\left(\int \sum_{a,b=1}^n \ve_a(z)\ve_b(z)\right)^i
\int\sum_{a=1}^n\si_a(z) d^2z = C_i \int\sum_{a=1}^n\si_a(z) d^2z
$$
We give here directly the result (see eq.(\ref{rgeh}) in appendix C)~:
\bea
r^{-{15\over8}-a(\epsilon)}h(r) &=& h_0 ( 1 + (n-1)g_0^2 \pi^2
{r^{-6\epsilon}\over 2}\left[1 + {4\over3}(2-\lambda)
{\Gamma^2(-{2\over3})\Gamma^2({1\over6})\over\Gamma^2(-{1\over3})
\Gamma^2(-{1\over6})}\right] \\
 &&-12(n-1)(n-2) g_0^3 \pi^3
\left(r^{-9\epsilon}\over9\epsilon\right)\left[1 + {8\over9}(2-\lambda)
{\Gamma^2(-{2\over3})\Gamma^2({1\over6})\over\Gamma^2(-{1\over3})
\Gamma^2(-{1\over6})}\right]) \nn
\eea
The multiplicative term $r^{-{15\over8}-a(\epsilon)}$ in front of $h(r)$ is
introduced in order to make this parameter dimensionless. Here,
$a(\epsilon)$ is a function of $\epsilon$ depending on which representation
of the spin field we are taking in the Coulomb gas picture (see section
2). Its explicit form will be irrelevant in the following. The
corresponding R.G. equation for $Z_\si$ will be given by
\bea
\label{eqzsi}
{dln(Z_\si(r))\over dln(r)} =  -3(n-1)g^2(r) \pi^2 \epsilon
\left[1 + {4\over3}(2-\lambda)
{\Gamma^2(-{2\over3})\Gamma^2({1\over6})\over
\Gamma^2(-{1\over3})\Gamma^2(-{1\over6})}\right] \nn \\
+ 4 (n-1)(n-2) \pi^3 g^3(r)
\eea
\section{Correlation Functions}
We now have all the ingredients needed in order to compute the correlation
functions. They will be calculated with the help of the R.G. equations, for
the theory with $m_0$, $h_0\rightarrow 0$. From the R.G. equations, we
have~:
$$
<\varepsilon(0)\varepsilon(sR)>_{r,g(r)} =
{Z^2_\ve(sr,g(sr))\over Z^2_\ve(r,g(r))}s^{-2\Delta_{\varepsilon}}
<\varepsilon(0)\varepsilon(R)>_{r,g(sr)}
$$
This can be written as~:
\beq
\label{zeps}
<\varepsilon(0)\varepsilon(sR)>_{r,g(r)} =
e^{2\int\limits_{g_0}^{g(s)}{\gamma_\ve(g)\over \beta(g)}
dg}s^{-2\Delta_{\varepsilon}} <\varepsilon(0)\varepsilon(R)>_{r,g(sr)}
\eeq
where we used the notation~:
\beq
\label{defzve}
{dlnZ_\ve\over dlnr} = \gamma_\ve(g)
\eeq
and $g(s) = g(sr); g_0 = g(r)$. We assume now $r$ to be a lattice cut-off
scale. In a similar way for $<\sigma(0)\sigma(R)>$ the R.G. equation is~:
\beq
\label{zeps2}
<\sigma(0)\sigma(sR)>_{r,g(r)} =
e^{2\int\limits_{g_0}^{g(s)}{\gamma_\si(g)\over \beta(g)}
dg}s^{-2\Delta_{\sigma}} <\sigma(0)\sigma(R)>_{r,g(sr)}
\eeq
with
\beq
\label{defzsi}
{dlnZ_\si\over dlnr} = \gamma_\si(g)
\eeq
In equations (\ref{zeps})-(\ref{zeps2}), $R$ is an arbitrary scale
which can be fixed to one lattice spacing $r$ of a true statistical model.
The dependence of $<\sigma(0)\sigma(r)>_{r,g(s)}$ on $s$ will then be
negligible, assuming that there are not interactions on distances
smaller than $r$. Therefore, it reduces to a constant. Then, $s$ will
measure the number of lattice spacings between two spins in
$<\sigma(0)\sigma(sR)>$. In the following, when we treat separately
the Ising model and the 3-states Potts model, we adopt the choice $r=1$.
\subsection{The Ising model}
The Ising model corresponds to the case $\epsilon \rightarrow 0$ and so the
$\beta$ function is~:
\beq
\label{beti}
\beta(g) =  - 8 \pi g^2 + 32 \pi^2 g^3
\eeq
Therefore, we can see that the I.R. fixed point is located at $g=0$. Also
we have, by eqs.(\ref{eqzve}), (\ref{eqzsi}) for $n=0, \epsilon=0$ and
definitions (\ref{defzve}), (\ref{defzsi}),
\bea
\gamma_\varepsilon (g)&=&-4\pi g + 8\pi^2 g^2 \\
\gamma_\sigma (g)&=&8\pi g^3
\eea
The integral for the $\varepsilon$ correlation
function, eq.(\ref{zeps}), gives~:
\bea
2\int\limits_{g_0}^{g(s)}{\gamma_\ve(g)\over \beta(g)} dg &=&
\int\limits_{g_0}^{g(s)} {1-2\pi g\over 1- 4\pi g}{dg\over g}
\approx  \int\limits_{g_0}^{g(s)} (1+2 \pi g){dg\over g}\nn\\
&=& 2\pi(g(s)-g_0)+ ln\left(g(s)\over g_0 \right)
\eea
Now, we need to compute $g(s)$. The integration of equation
$\beta(g)={dg\over dln(r)}$ gives~:
$$
\int\limits_{g_0}^{g(s)} {dg\over -8\pi g^2 + 32 \pi^2 g^3} =
\int\limits_{r}^{sr} dlnr
$$
with the following solution up to the second order~:
\beq
g(s) = {g_0\over 1+8\pi g_0 ln(s)}(1+{4\pi g_0 ln(1+8\pi g_0 ln(s))\over
1+8\pi g_0 ln(s) })+ O(g_0^3)
\eeq
So, $<\varepsilon \varepsilon>$ correlation function is given by~:
\bea
&&<\varepsilon(0) \varepsilon(s)>_{g_0} \sim {g(s)\over g_0} \left( 1 -
2\pi(g_0 - g(s))\right) s^{-2\Delta_{\varepsilon}} \\
&\sim& {1\over 1+8\pi g_0 ln(s)}\left(1+{4\pi g_0\over 1+8\pi g_0 ln(s)}(
ln(1+8\pi g_0 ln(s)) - 4\pi g_0 ln(s))\right) s^{-2\Delta_{\ve}} + O(g_0^2)
\nn
\eea
For the $\sigma$ correlation function, the computation is similar. In fact,
in that case, we have $\gamma_\si(g)=8\pi^3 g^3$. Thus, keeping only the
first order of the $\beta$-function (\ie $\beta(g) =  - 8 \pi g^2$), we
obtain~:
\beq
2\int\limits_{g_0}^{g(s)}{\gamma_\si(g)\over \beta(g)} dg =
-2\pi^2 \int\limits_{g_0}^{g(s)}gdg
= -\pi^2 \left( g(s)^2 - g_0^2 \right) + O(g_0^3)
\eeq
The $<\sigma \sigma>$ correlation function is then found to be given by~:
\bea
<\sigma(0) \sigma(s)> &\sim& \left( 1 + \pi^2  \left( g_0^2
- g(s)^2 \right) \right)  s^{-2\Delta_{\sigma}} \\
&\sim&
 \left( 1 + \pi^2 g_0^2  \left( 1 - {1\over (1+8\pi g_0 ln(s)
)^2 }\right)\right)
 s^{-2\Delta_{\sigma}}+O(g_0^3) \nn
\eea
The calculation of the $g^3$ term in the $\beta$ function and the $g^2$
term in the renormalisation of $\varepsilon$ was already done in
\cite{ludwig1}, extending the one loop result of \cite{dots1}. We recovered
these higher order corrections using a different technique, which allowed
us to calculate also the modified correlation function of the spin
operators.

\subsection{The Potts model}
We now consider the 3-state Potts model. With the convention that we use
this case corresponds to $\epsilon = -{2\over15}$. $\beta(g)$ is given in
eq.(\ref{tobf}), and
\bea
\gamma_\varepsilon (g) &=& -4\pi g + 8\pi^2 g^2 \\
\gamma_\sigma (g) &=& 3\pi ^2 \epsilon \left( 1 +{4\over 3}(2-\lambda)
{\Gamma^2(-{2\over 3}) \Gamma^2({1\over 6}) \over \Gamma^2(-{1\over 3})
\Gamma^2 (-{1\over 6})}\right) g^2 + 8\pi^3 g^3
\eea
At long distances, the integrals in eqs.(\ref{zeps}),
(\ref{zeps2}) will be dominated by the
region $g \sim g_c$, with $g_c = -{3\epsilon \over 8
\pi} + {9\epsilon^2 \over 16\pi} + O(\epsilon^3) $. This is
different from the Ising model, because here
$\gamma_\ve(g_c)$ and $ \gamma_\si(g_c)$ have finite values for
$g=g_c$. Thus,
\beq
\int\limits_{g_0}^{g(s)}{\gamma_\ve(g)\over \beta(g)} dg \approx
\gamma_\ve(g_c) ln(s)
\qquad\mbox{and}\qquad
\int\limits_{g_0}^{g(s)}{\gamma_\sigma(g)\over \beta(g)} dg \approx
\gamma_\ve(g_c) ln(s)
\eeq
The correlation functions can then be deduced directly~:
\beq
<\varepsilon(0) \varepsilon(s)>_{g_0} \sim
s^{-(2\Delta_{\varepsilon}-2\gamma_\ve(g_c)) }
\quad \mbox{and} \quad
<\sigma(0) \sigma(s)>_{g_0} \sim
s^{-(2\Delta_{\sigma}-2\gamma_\si(g_c)) }
\eeq
So, we can see that a direct consequence of the new IR fixed point is a
modification of the critical exponents $\Delta_{\varepsilon}$ and
$\Delta_{\sigma}$. A straightforward computation will give these new
exponents~:
\bea
2\Delta'_{\varepsilon} = 2\Delta_{\varepsilon} - 2\gamma_\ve(g_c) &=&
2\Delta_{\varepsilon} + 8\pi g_c - 16 \pi^2 g_c^2\nn\\
 &=& 2\Delta_{\varepsilon} -3\epsilon
+{9\over4}\epsilon^2 + O(\epsilon^3)
\eea
and
\bea
2\Delta'_{\sigma} = 2\Delta_{\sigma} - 2\gamma_\si(g_c)  &=&
 2\Delta_{\sigma} - 6 \pi^2 g_c^2 \epsilon \left[1 + {4\over3}(2-\lambda)
{\Gamma^2(-{2\over3})\Gamma^2({1\over6})\over\Gamma^2(-{1\over3})
\Gamma^2(-{1\over6})}\right] - 16 \pi^3 g_c^3 \nn\\
&=& 2\Delta_{\sigma} - {9\over8} (2-\lambda)
{\Gamma^2(-{2\over3})\Gamma^2({1\over6})\over\Gamma^2(-{1\over3})
\Gamma^2(-{1\over6})} \epsilon^3 + O(\epsilon^4)
\eea
Here $\lambda = {1\over2}$ and the final result for the new critical
exponent is~:
\beq
\label{ndim}
2\Delta'_{\sigma} = 2\Delta_{\sigma} - {27\over16}
{\Gamma^2(-{2\over3})\Gamma^2({1\over6})\over\Gamma^2(-{1\over3})
\Gamma^2(-{1\over6})} \epsilon^3 + O(\epsilon^4)
\eeq
The value of the critical exponent for the energy operator was already
computed, up to the second order, by Ludwig \cite{ludwig1}. For the spin
operator, deviation of the critical exponent from the pure case appears
only at the third order, which we compute. This deviation is in fact very
small. While in case of 3-state Potts model without disorder
$2\Delta_{\sigma} = {4\over 15}$, we obtain the new  critical
exponent $2\Delta'_{\sigma} = {4\over 15} + 0,00264 = 0,26931$. Thus, the
deviation corresponds to a modification of $1\%$.
\section{Conclusions}
In case of Ising model spin-spin function the calculation of up to third
order of the renormalisation group was needed to find the deviation from
the perfect model case, in the form of the cross-over in the
amplitude. This completes the observations of
\cite{shalaev,shankar,ludwig2}, based on absence of renormalisation of this
function in the first order, that asymptotically the spin-spin function has
the same exponent as in case of the perfect lattice model.

Earlier calculation of the spin-spin function in the second reference of
\cite{dots1}, in which a non-local fermionic representation had been used,
gave a different result, which was not confirmed neither by further
theoretical calculations, based on local renormalisation directly of spin
operators, nor by the numerical simulations in \cite{adsw,talapov}. In
calculation of \cite{dots1} a double summation series had been involved,
both sums being log divergent. One sum was due to non-local representation
of the spin-spin function, another was due to randomness-caused
interactions. In the calculations, a particular ordering of the two
summations had been employed : renormalizing the terms of the first series
and summing up next. We may guess, lacking a better established reason,
that such a treatment might not be justified.

Recently the numerical simulations of the random Ising model has been
performed which measure directly  the deviation of $<\sigma \sigma> $ from
the pure Ising model at the critical point \cite{talapov}. These
measurements were made for disorder such that
$8\pi g_0 \approx 0.3$. Deviations predicted by our
computations are very small. They correspond to $0.1 \%$. The deviations
obtained in numerical simulations are around ten times larger, and they are
of opposite sign, \ie $ $ correspond to an extra decrease of the spin-spin
function
with distance $r$. In \cite{talapov}, it has been checked that this
decrease corresponds, within the accuracy of the measurements, to a factor
function of the ratio ${r/L}$, $F({r/L})$, $r$ being the distance between
the spins and $L$ is the lattice size. So they correspond to finite size
effects, being different for perfect and random models. We would suggest,
on the bases of our calculation of the $r$ dependence of the
spin-spin function on an infinite lattice, that numerical deviations will
continue to be plotted by the same curve $F({r/L})$, if one measures
$<\sigma \sigma>$ for different lattice sizes as it has been done in
\cite{talapov}, until the accuracy reaches the value of the $r$-deviation
which we calculated here. Only then the curves for different $L$ will
split.

In case of Potts model, the deviation of $<\sigma \sigma>$
appear also only in the third order like for the Ising model, this time in
the exponent of the asymptotic
function, eq.(\ref{ndim}). This might be easier to check with numerical
simulations \cite{usf}. We refered to the 3-state Potts model in Sec.4.2
because this particular case might be easier to realize numerically,
and also because we expect that $\epsilon$-expansion should be well defined
in this case. Extending our result to the 4-state Potts model might
be questionable, as this is actually the limiting case in the range of
minimal conformal theories which are parametrized by $\epsilon$.

\noindent{\large\bf Acknowledgements}

We are grateful for A.~L.~Talapov and L.~N.~Shchur for keeping us informed
prior to publication on their numerical simulation results, which are of
very high accuracy and for the first time obtained directly for the
spin-spin function, in coordinate space. Very useful and stimulating
discussions with D.~Bernard on the analytic approach and the calculations
presented in this paper are gratefully acknowledged.

\appendix
\section{Renormalisation of $g$}
In this appendix, we shall describe the calculations for the
renormalisation of $g$ up to the second order.
\subsection{First order}
The first order correction to $g$ comes from the contraction of two $
\left(\varepsilon \varepsilon \right) $ terms (plus some combinatorial
factor corresponding to all of the possible contractions) :
$$
{g_0^2\over2} \int\limits_{|x-y|<r} \displaystyle\sum_{a\not= b}
\varepsilon_a(x) \varepsilon_b(x)  \displaystyle\sum_{c\not= d}
\varepsilon_c(y) \varepsilon_d(y) d^2x d^2y
$$
When $b=c\not= a, d$ this contractions  gives~:
$$
 A_2(r,\epsilon) g_0^2 \int \displaystyle\sum_{a\not= d}
\varepsilon_a(x) \varepsilon_d(x) d^2x
$$
Here, $A_2(r,\epsilon)$ is the result of the integral produced by the
contraction of two $\varepsilon$ operators~:
$$
 A_2(r,\epsilon) = 2(n-2) \int\limits_{|y-x|<r}
<\varepsilon(x)\varepsilon(y)>_0 d^2y = 4\pi (n-2)
\int\limits_{y<r}\left(dy\over y^{1+3\epsilon}\right)
$$
\beq
 = -4\pi (n-2)\left(r^{-3\epsilon}\over 3\epsilon\right)
\eeq
$<\varepsilon(x)\varepsilon(y)>_0$ being the standard correlation function
of the unperturbed conformal field theory.
\subsection{Second order}
We will now compute the second order corrections to $g$, produced by the
contractions of three $\left(\varepsilon \varepsilon \right)$ terms. Here,
there will be two contributions coming from~:
\beq
\label{2nn}
{g_0^3\over3!} \int\limits_{|x-y|,|x-z|<r} \displaystyle\sum_{a\not= b}
\varepsilon_a(x) \varepsilon_b(x)  \displaystyle\sum_{c\not= d}
\varepsilon_c(y) \varepsilon_d(y) \displaystyle\sum_{e\not= f}
\varepsilon_e(z) \varepsilon_f(z) d^2x d^2y d^2z
\eeq
The first one corresponds to the case when $ b = c ; d = e ; b,d \not=
a,f$. We will denote this term by~:
$$
 A_{3,1}(r,\epsilon) g_0^3 \int \displaystyle\sum_{a\not= f}
\varepsilon_a(x) \varepsilon_f(x) d^2x
$$
where $A_{3,1}(r,\epsilon)$ corresponds to the following~:
\beq
\label{ee}
 A_{3,1}(r,\epsilon) = 4(n-2)(n-3) \int\limits_{|y-x|,|z-x|<r}
<\varepsilon(x)\varepsilon(y)>_0 <\varepsilon(y)\varepsilon(z)>_0 d^2z d^2y
\eeq
Now, replacing $<\varepsilon(x)\varepsilon(y)>_0 $ by
$|x-y|^{-2-3\epsilon}$ and performing a trivial change of variable, this
gives~:
\bea
\label{atu}
A_{3,1}(r,\epsilon) &=&  8\pi (n-2)(n-3)
\int\limits_{y<r}\left(dy\over y^{1+6\epsilon}\right) \int
|z|^{-2-3\epsilon} |z-1|^{-2-3\epsilon} d^2z \nn\\
&= & 16\pi^2 (n-2)(n-3)\left(r^{-6\epsilon}\over
9\epsilon^2\right)
\eea
The second term is produced when $a = c = e \not= f; b = d \not= f$. This
will be denoted by~:
$$
A_{3,2}(r,\epsilon) g_0^3 \int \displaystyle\sum_{a\not= f}
\varepsilon_a(x) \varepsilon_f(x) d^2x
$$
with:
\beq
\label{gg}
 A_{3,2}(r,\epsilon) = 4 (n-2)  \int\limits_{|y-x|,|z-x|<r}
<\varepsilon(x) \varepsilon(y) \varepsilon(z) \varepsilon(\infty)>_0 <
\varepsilon(y)  \varepsilon(z)>_0 d^2y d^2z
\eeq
Here, the $<\varepsilon(x) \varepsilon(y) \varepsilon(z)
\varepsilon(\infty)>_0$ term corresponds to the result of $ \varepsilon
\varepsilon \varepsilon \rightarrow \varepsilon$ obtained by projecting
$\varepsilon \varepsilon \varepsilon $ over $\varepsilon (\infty)$.
In order to compute $A_{3,2}(r,\epsilon)$, we need to use the Coulomb gas
representation for this $4-$points correlation function~:
\bea
A_{3,2}(r,\epsilon) &=& 4 (n-2) N \times \nn \\
 \int\limits_{|y-x|,|z-x|<r}& & \!\!\!\!\!\!\!\!\!\!\!\!\!\!\!
<V_{\overline{12}}(x) V_{12}(y) V_{12}(z) V_+(u) V_{12}(\infty)>
|y-z|^{-2-3\epsilon} d^2y d^2z d^2u
\eea
$N$ is a normalization constant which can be fixed by the OA
relation in the following way. We have~:
$$
<\varepsilon(0) \varepsilon(R) \varepsilon(x) \varepsilon(y)> =  N \int
<V_{\overline{12}}(0) V_{12}(R) V_{12}(x) V_+(u) V_{12}(y)> d^2u
$$
In the limit $ R \rightarrow 0 $~:
$$
\varepsilon(0) \varepsilon(R) = {1\over R^{4 \Delta_{12}}} I + ...
$$
( $I$ is the identity operator) and
$$
<\varepsilon(0) \varepsilon(R) \varepsilon(x) \varepsilon(y)> \approx
{1\over R^{4 \Delta_{12}}} <\varepsilon(x) \varepsilon(y)> = {1\over R^{4
\Delta_{12}} |x-y| ^{4 \Delta_{12}}}
$$
For the r.h.s. we obtain
$$
N {1\over R^{4 \Delta_{12}}} \int <V_{2 \alpha_0}(0) V_{12}(x) V_{12}(y)
V_+(u)> d^2u
$$
which produces
$$
N  {1\over R^{4 \Delta_{12}} |x-y| ^{4 \Delta_{12}}} \int |u|^{4
\alpha_{12} \alpha_+} |u-1|^{8 \alpha_0 \alpha_+} d^2u
$$
Comparing both results we obtain~:
\beq
\label{n}
N = \Bigl( \int |u|^{4
\alpha_{12} \alpha_+} |u-1|^{8 \alpha_0 \alpha_+} d^2u \Bigr)^{-1} = -{2
\over \sqrt{3}} {(\Gamma(-{2 \over 3}))^2 \over(\Gamma(-{1 \over 3}))^4}
\eeq
We now return to the computation of $A_{3,2}(r,\epsilon)$. By redefining $
y \rightarrow y-x; z \rightarrow z-x$ , $A_{3,2}(r,\epsilon)$ is
transformed into the following integral
\bea
\label{gint}
A_{3,2}(r,\epsilon) &=& 8 \pi (n-2) N \times \\
&&\int\limits_{y<r}\left(dy\over y^{1+6\epsilon}\right)
\int |z|^{-4\Delta_{12}}  |z-1|^{-4\Delta_{12}+4\alpha^2_{12}}
|u|^{4\alpha_+ \alpha_{\overline{12}}} |u-1|^{4\alpha_+ \alpha_{12}}
|u-z|^{4\alpha_+ \alpha_{12}} d^2z d^2u \nn
\eea
The first integral equals $-{r^{-6\epsilon}\over 6\epsilon}$. The
second integral is more complicated. We will show in appendix D how to
calculate it. In fact, it equals (cf. (\ref{apa}))
$\sqrt{3}\pi{\Gamma^4(-{1\over3})\over
\Gamma^2(-{2\over 3})} $. We obtain~:
\beq
\label{atrde}
A_{3,2}(r,\epsilon) = 8 \pi^2 (n-2) \left(r^{-6\epsilon}\over
3\epsilon\right)
\eeq
In fact, in appendix D, the $z$ integration is performed over the whole
complex plane, while in the integral (\ref{gint}), the domain of
integration is restricted to the disk $|z|<{r\over |y|}$.
This introduces a new singularity at infinity, which we need to
subtract. However, we will show that this singularity is cancelled by
another singularity at the origin, in the computation that we
perform in appendix D. Thus, we need only to compute the singularity at the
origin, and this singularity must be added to the result. That the
singularity at the origin cancels the one at infinity is obvious if we
isolate these two parts in
$$
\int <\varepsilon(0) \varepsilon(1) \varepsilon(z) \varepsilon(\infty)> <
\varepsilon(1)  \varepsilon(z)> d^2z
$$
The singularity in $z\rightarrow 0$ is $ \int\limits_{|z|<r/|y|}d^2z
|z|^{-2-3\epsilon} = 2\pi{(r/|y|)^{-3\epsilon}\over -3\epsilon}$ and the
one in
$z\rightarrow \infty$ is $ \int\limits_{|z|>r/|y|}d^2z |z|^{-2-3\epsilon} =
2\pi{(r/|y|)^{-3\epsilon}\over 3\epsilon}$, confirming that these
contributions cancel each other.
We still need to compute the extra contribution from $z\rightarrow 0$.
Some easy manipulations show that this contribution is the same as the one
from the integral in equation (\ref{ee}). So the third second order term
is~:
\beq
\label{atd}
A_{3,3}(r,\epsilon) =  16\pi^2 (n-2)\left(r^{-6\epsilon}\over
9\epsilon^2\right)
\eeq
Finally, in the case $a=c=e$ and $b=d=f$, a straightforward calculation of
the integral (\ref{2nn}) gives us a result of the form
$\int\limits_{|z|<r}d^2z |z|^{-4-6\epsilon}$. This does not contain any
singularities. Collecting all of our results together, we obtain the
re-normalized coupling constant~:
\bea
 g(r) &=& g_0 + g_0^2 A_2(r,\epsilon) +
g_0^3 (A_{3,1}(r,\epsilon) + A_{3,2}(r,\epsilon) +
A_{3,3}(r,\epsilon))\nn\\
& =&g_0 \left(1  - g_0  4\pi (n-2){r^{-3\epsilon}\over
3\epsilon} + g_0^2 8 \pi^2 (n-2) {r^{-6\epsilon}\over
3\epsilon} \left(1+ {2 (n-2)\over 3\epsilon}\right)\right)
\eea
\section{Renormalisation of $m$}
This appendix is devoted to the computation of the renormalisation of the
coupling constant $m$ associated to the energy operator. As in appendix A,
we will
compute perturbatively by contracting $\varepsilon$ operators.

\subsection{First order}
The first order correction to $m$ is produced by the product of an
$g_0(\varepsilon \varepsilon)$ operator with an $m_0(\varepsilon )$
operator~:
$$
g_0 m_0 \int\limits_{|x-y|<r} \displaystyle\sum_{a\not= b}
\varepsilon_a(x) \varepsilon_b(x)  \displaystyle\sum_{c}
\varepsilon_c(y) d^2x d^2y
$$
When  $ b=c$, we can contract two $\varepsilon $ operators~:
$$
g_0 m_0 B_1(r,\epsilon) \int\displaystyle\sum_{a} \varepsilon_a(y) d^2y
$$
with
$$
 B_1(r,\epsilon) = 2(n-1) \int\limits_{|y-x|<r}
<\varepsilon(x)\varepsilon(y)>_0 d^2y
$$
\beq
 = 4\pi (n-1)
\int\limits_{y<r}\left(dy\over y^{1+3\epsilon}\right) = -4\pi
(n-1)\left(r^{-3\epsilon}\over
3\epsilon\right)
\eeq
\subsection{Second  order}
At the second order, corresponding to the product of two
$g_0(\varepsilon \varepsilon)$ operators with an $m_0(\varepsilon )$ one,
we have to compute~:
\beq
\label{2n}
{g_0^2\over2} m_0 \int\limits_{|x-y|,|x-z|<r} \displaystyle\sum_{a\not= b}
\varepsilon_a(x) \varepsilon_b(x)  \displaystyle\sum_{c\not= d}
\varepsilon_c(y) \varepsilon_d(y) \displaystyle\sum_{e}
\varepsilon_e(z)  d^2x d^2y d^2z
\eeq
This expression is, in fact, very similar to the one in
eq.(\ref{2nn}). Again there will be two contributions. The first one
corresponds to $ b=c \not= d=e $. We will denote this case by~:
$$
g_0^2 m_0 B_{2,1}(r,\epsilon) \int\displaystyle\sum_{a} \varepsilon_a(y)
d^2y
$$
$B_{2,1}(r,\epsilon)$ corresponds to $A_{3,1}(r,\epsilon)$ of
appendix A, where the combinatorial term $(n-2)(n-3)$ has been replaced by
$(n-1)(n-2)$. From eq.(\ref{atu}) we can read the result~:
\beq
B_{2,1}(r,\epsilon) =  16\pi^2 (n-1)(n-2)\left(r^{-6\epsilon}\over
9\epsilon^2\right)
\eeq
The second case is produced when $b=c=e \not= a=d$~:
$$
g_0^2 m_0 B_{2,2}(r,\epsilon) \int\displaystyle\sum_{a} \varepsilon_a(y)
d^2y
$$
Again, $B_{2,2}(r,\epsilon)$ corresponds to
$A_{3,2}(r,\epsilon)$ computed in appendix A where $(n-2)$ has been
replaced by ${(n-1)\over 2}$. From eq.(\ref{atrde}), we have
\beq
B_{2,2}(r,\epsilon) = 4 \pi^2 (n-1) \left(r^{-6\epsilon}\over
3\epsilon\right)
\eeq
Again, the $B_{2,2}(r,\epsilon)$ term must be completed by a
$B_{2,3}(r,\epsilon)$ term (the equivalent of $A_{2,3}(r,\epsilon)$) which
is~:
\beq
B_{2,3}(r, \epsilon) =  8\pi^2 (n-1)\left(r^{-6\epsilon}\over
9\epsilon^2\right)
\eeq
Collecting all of these results together, we then obtain for $m(r)=m_0
\Bigl( 1+B_1(r, \epsilon) g_0 + g_0^2 (B_{2,1}(r, \epsilon) +  B_{2,2}(r,
\epsilon) + B_{2,3}(r, \epsilon) \Bigr)$ at the second order
in $g(r)$~:
\beq
 m(r) = m_0 \Bigl( 1-4\pi
(n-1)g_0 {r^{-3\epsilon}\over
3\epsilon} + 4 \pi^2 (n-1)g_0^2{r^{-6\epsilon}\over
3\epsilon} (1+{4n-6\over 3\epsilon} )
\Bigr)
\eeq
\section{Renormalisation of $h$}
We now turn to the computation of the renormalisation of the coupling
constant $h(r)$ associated to the $\sigma$ operator. Here the situation
is a little bit different, because
the $\varepsilon$ operator {\it and} the $\sigma$ operator are involved. At
the first order, we must compute the product of $g_0(\varepsilon
\varepsilon)$ on $h_0(\sigma)$. This product will give in fact no
contribution (the OPA of $\sigma$ and $\varepsilon$ does not contain
$\sigma$ operators at the first order) and so we go directly to the second
order.
\subsection{Second order}
Here, we will compute the product of two $g_0(\varepsilon \varepsilon)$
with an
$h_0 (\sigma)$ operator~:
$$
{g_0^2\over2} h_0 \int\limits_{|x-y|,|x-z|<r} \displaystyle\sum_{a\not= b}
\varepsilon_a(x) \varepsilon_b(x)  \displaystyle\sum_{c\not= d}
\varepsilon_c(y) \varepsilon_d(y) \displaystyle\sum_{e}
\sigma_e(z)  d^2x d^2y d^2z
$$
When $ a=c=e\not=b=d$ the product  $\sigma\varepsilon\varepsilon$ will
contain a $\sigma$ operator. By projecting $\sigma\varepsilon\varepsilon$
on a $\sigma(\infty)$ operator, we obtain~:
$$
 h_0  g_0^2 C_2(r,\epsilon)\int\displaystyle\sum_{a=1}^n \sigma_a(z) d^2z
$$
where
\beq
C_2(r,\epsilon) = 2 (n-1)  \int\limits_{|y-x|,|z-x|<r}
<\sigma(x) \varepsilon(y) \varepsilon(z) \sigma(\infty)>_0 <
\varepsilon(y)  \varepsilon(z)>_0 d^2y d^2z
\eeq
Using the Coulomb gas representation, this integral becomes~:
\bea
\label{c2}
& & 4 (n-1)N \pi \int\limits_{y<r}\left(dy\over y^{1+6\epsilon}\right)
\times\\
& & \int |z|^{4\alpha_{12}\alpha_{\overline{k,k-1}}}
|z-1|^{-4\Delta_{12}+4\alpha^2_{12}}
|u|^{4\alpha_+ \alpha_{\overline{k,k-1}}} |u-1|^{4\alpha_+ \alpha_{12}}
|u-z|^{4\alpha_+ \alpha_{12}} d^2z d^2u \nn
\eea
where the normalisation $N$ is the same as the one computed in the appendix
A, see eq.(\ref{n}). The first part of this expression (the $y$ integral)
gives ($-{r^{-6\epsilon}\over 6\epsilon}$), while the second one is
computed in appendix D, see (\ref{cc2}). We obtain~;
\beq
\label{x}
C_2(r,\epsilon) = (n-1) \pi^2
{r^{-6\epsilon}\over 2}\left[1 + {4\over3}(2-\lambda)
{\Gamma^2(-{2\over3})\Gamma^2({1\over6})\over\Gamma^2(-{1\over3})
\Gamma^2(-{1\over6})}\right]
\eeq
\subsection{Third order}
At the third order, corresponding to the product of three $g_0(\varepsilon
\varepsilon)$ operators with a $h_0(\sigma)$ one, we have to compute~:
$$
{g_0^3\over3!} h_0 \int\limits_{|x-y|,|x-z|,|x-u| <r}
\displaystyle\sum_{a\not= b}
\varepsilon_a(x) \varepsilon_b(x)  \displaystyle\sum_{c\not= d}
\varepsilon_c(y) \varepsilon_d(y)
 \displaystyle\sum_{e\not= f}\varepsilon_e(x) \varepsilon_f(x)
\displaystyle\sum_{g}\sigma_g(z)  d^2x d^2y d^2z d^2u
$$
Here contributions will come from the following contractions
$a=c=g ; b=e ; d=f$ and produce~:
$$
 h_0  g_0^3 C_3(r,\epsilon)\int\displaystyle\sum_{a=1}^n \sigma_a(z) d^2z
$$
where
\bea
 C_3(r,\epsilon) &=& 4(n-1)(n-2) \times \\
 \int\limits_{|y-x|,|z-x|,|w-x| <r}&&\!\!\!\!\!\!\!\!\!\!\!\!\!\!
<\sigma(x) \varepsilon(y) \varepsilon(z) \sigma(\infty)>_0
<\varepsilon(y)  \varepsilon(w)>_0<\varepsilon(w)  \varepsilon(z)>_0 d^2y
d^2z d^2w \nn
\eea
Again, this expression can be computed in the Coulomb gas representation
$$
 8(n-1)(n-2)N \pi \int
|w|^{-2-3\epsilon} |w-1|^{-2-3\epsilon} d^2w\int\limits_{y<r}\left (dy\over
y^{1+9\epsilon}\right)
$$
\beq
\label{c3}
\int |z|^{4\alpha_{12}\alpha_{\overline{k,k-1}}}
|z-1|^{-8\Delta_{12}+4\alpha^2_{12}+2}
|u|^{4\alpha_+ \alpha_{\overline{k,k-1}}} |u-1|^{4\alpha_+ \alpha_{12}}
|u-z|^{4\alpha_+ \alpha_{12}} d^2z d^2u
\eeq
where we also factorize the $y$ and $w$ integrations. The first two
integrations will then produce ${24\pi r^{-9\epsilon} \over 81 \epsilon^2}
$ while the results of the $z$ and $u$
integrations are given by eq.(\ref{cc7}) of appendix D. So~:
\beq
C_3(r,\epsilon) = -12(n-1)(n-2) \pi^3
\left(r^{-9\epsilon}\over9\epsilon\right)\left[1 + {8\over9}(2-\lambda)
{\Gamma^2(-{2\over3})\Gamma^2({1\over6})\over\Gamma^2(-{1\over3})
\Gamma^2(-{1\over6})}\right]
\eeq
Putting together the $C_i$, we obtain for $h(r)$, up to the third order~:
\bea
\label{rgeh}
h(r) &=& h_0 ( 1 + (n-1)g_0^2 \pi^2
{r^{-6\epsilon}\over 2}\left[1 + {4\over3}(2-\lambda)
{\Gamma^2(-{2\over3})\Gamma^2({1\over6})\over\Gamma^2(-{1\over3})
\Gamma^2(-{1\over6})}\right] \\
 &&-12(n-1)(n-2) g_0^3 \pi^3
\left(r^{-9\epsilon}\over9\epsilon\right)\left[1 + {8\over9}(2-\lambda)
{\Gamma^2(-{2\over3})\Gamma^2({1\over6})\over\Gamma^2(-{1\over3})
\Gamma^2(-{1\over6})}\right]) \nn
\eea

\section{Computation of integrals}
In this appendix, we briefly present the general method to compute
integrals of the following form
\beq
I =
\int |x|^{2a}  |x-1|^{2b}
|y|^{2a'} |y-1|^{2b'}
|x-y|^{4g} d^2x d^2y
\eeq
where the integration is performed over the whole complex plane.
Using the techniques of ref.\cite{dots2} (see also \cite{lectvol}), this
integral can be  decomposed into holomorphic and antiholomorphic parts~:
\beq
\label{i}
I = s(b) s(b') \left[ J_1^+  J_1^- + J_2^+  J_2^-\right]
+ s(b) s(b'+2g) J_1^+ J_2^- + s(b+2g) s(b') J_2^+ J_1^-
\eeq
where $s(x)$ corresponds to $sin(\pi x)$ and
\bea
\label{jis}
J_1^+ &=& J(a,b,a',b',g)~;~J_2^+ = J(b,a,b',a',g) \nn \\
J_1^- &=& J(b,-2-a-b-2g,b',-2-a'-b'-2g,g) \\
J_2^- &=& J(-2-a-b-2g,b,-2-a'-b'-2g,b',g) \nn
\eea
Here, we used the notation
\bea
\label{hyp}
 J(a,b,a',b',g) =
\int\limits_{0}^{1} du \int\limits_{0}^{1} dv~
u^{a+a'+2g+1} (1-u)^b v^{a'}&&\!\!\!\!\!\!\!\! (1-v)^{2g} (1-uv)^{b'}\\
= {\Gamma(2+a+a'+2g)\Gamma(1+b)\Gamma(1+a')\Gamma(1+2g)\over
\Gamma(3+a+a'+b++2g)\Gamma(2+a'+2g)}
&&\!\!\!\!\!\!\!\!\sum_{k=0}^{\infty}{(-b')_k (2+a+a'+2g)_k (1+a')_k \over
k! (3+a+a'+b++2g)_k (2+a'+2g)_k} \nn
\eea
and
$$
(a)_k = a(a+1)...(a+k-1)
$$
The $J$ integrals appearing in (\ref{i}) are not all independent. Using
contour deformation of integrals it can be shown that we have the following
relations:
\beq
\label{rel1}
s(2g+a+b)J_1^- + s(a+b)J_2^- = {s(a)\over s(2g+a'+b')}
\left( s(a') J_1^+ + s(2g+a')J_2^+ \right)
\eeq
\beq
\label{rel}
s(2g+a'+b')J_2^- + s(a'+b')J_1^- = {s(a')\over s(2g+a+b)}
\left( s(a) J_2^+ + s(2g+a)J_1^+ \right)
\eeq
With these formulas, the integrals appearing in (\ref{gint}), (\ref{c2}),
(\ref{c3}) can be calculated as a power series of $\epsilon$. We will now
compute explicitly these three cases~:
\begin{itemize}
\item We first compute the integral (\ref{gint}). The coefficients
corresponding to that case are
\beq
a=-2a' =2+3\epsilon \quad ; \quad b'=-{1\over 3} -\epsilon \quad ; \quad
b=-2g=-{2\over 3} -{\epsilon \over 2}
\eeq
Substituting these values in (\ref{i}), we obtain at the first order in
$\epsilon$
\beq
I={3\over4}(J_2^+ -J_1^+)(J_1^- - J_2^-) -{\sqrt{3}\pi\epsilon\over
4}\left( J_1^+(2J_1^- +J_2^-)+J_2^+(J_1^- +2J_2^-)\right)
\eeq
Now writing the $J_i^-$ as functions of the $J_i^+$, with the help of
relations (\ref{rel1},\ref{rel}), this simplifies to
\beq
I=3\sqrt{3}\pi\epsilon J_1^+ J_2^+ -{9\over 2}\pi^2\epsilon^2 (J_2^+)^2
\eeq
The $J_1^+$ and $J_2^+$ can now be computed explicitly using formulas
(\ref{jis},\ref{hyp}). We obtain~:
\bea
J_1^+ &=&{2\over 9} ({3\over 2}+\pi\sqrt{3}){\Gamma^2(-{1\over3})\over
\Gamma({1\over 3})} + O(\epsilon) \nn\\
J_2^+ &=&{4\over 9\epsilon}{\Gamma^2(-{1\over3})\over
\Gamma({1\over 3})} + O(cst) \nn
\eea
Collecting all of these pieces together, we obtain, at the lowest order in
$\epsilon$~:
\beq
\label{apa}
I= \sqrt{3}\pi{\Gamma^4(-{1\over3})\over
\Gamma^2(-{2\over 3})}
\eeq
\item We now compute the integral (\ref{c2}). Here, we have~:
\beq
a=-{1\over3} + \lambda\epsilon =-{1\over2} a' \quad ; \quad
b = 2g = b'-1 = -{4\over3} - \epsilon
\eeq
Then, relating the $J_i^-$ to the $J_i^+$, we obtain for $I$~:
\beq
\label{i2}
I = {3^{3\over2} \pi \epsilon\over 2}\left(J_2^-\right)^2 +
 {3^{1\over2} (2-\lambda)\pi \epsilon\over 2}\left(J_2^- -J_1^- \right)^2
+ O(\epsilon^2)
\eeq
These $J_i^-$ can again be computed with the help of
(\ref{jis},\ref{hyp})~:
$$
J_1^- = J_2^- + {\Gamma(-{1\over3})\Gamma({1\over6})\over
\Gamma(-{1\over6})} = {1\over2} {\Gamma^2(-{1\over3})\over
\Gamma(-{2\over3})} +  {\Gamma(-{1\over3})\Gamma({1\over6})\over
\Gamma(-{1\over6})}+ O(\epsilon)
$$
Thus~:
\beq
\label{cc2}
I = {3^{3\over2} \pi \epsilon\over 8} {\Gamma^4(-{1\over3})\over
\Gamma^2(-{2\over3})} + {3^{1\over2} (2-\lambda)\pi \epsilon\over 2}
{\Gamma^2(-{1\over3})\Gamma^2({1\over6})\over
\Gamma^2(-{1\over6})}+ O(\epsilon^2)
\eeq
\item The last case, corresponding to (\ref{c3}), use the following
parameters
\beq
a=-{1\over3} + \lambda\epsilon =-{1\over2} a' \quad ; \quad
b = 2g =  -{4\over3} - \epsilon \quad ; \quad
b'=-{1\over3}-{5\over2}\epsilon
\eeq
Performing the same manipulations, we immediately obtain
\beq
\label{cc7}
I = {9\sqrt{3} \pi \epsilon\over 16} {\Gamma^4(-{1\over3})\over
\Gamma^2(-{2\over3})} + {3^{1\over2} (2-\lambda)\pi \epsilon\over 2}
{\Gamma^2(-{1\over3})\Gamma^2({1\over6})\over
\Gamma^2(-{1\over6})}+ O(\epsilon^2)
\eeq

\end{itemize}
%%%%% ***************************************** %%%%%%%%

\end{document}